\documentclass[12pt]{article}
\usepackage[utf8]{inputenc}
\pdfoutput=0
\usepackage[lmargin=1in,rmargin=1in,tmargin=35mm,bmargin=35mm]{geometry}
\usepackage{setspace}

\def\bSig\mathbf{\Sigma}

\usepackage{amsmath}
\newcommand\numberthis{\addtocounter{equation}{1}\tag{\theequation}}

\usepackage{booktabs}
\usepackage{graphicx}

\makeatletter
\renewcommand\@biblabel[1]{}
\makeatother

\usepackage{booktabs}

\usepackage{array}

\usepackage{authblk}

\usepackage[small]{titlesec}

\titleformat{\section}
  {\normalfont\fontsize{12}{12}\bfseries}{\thesection}{1em}{}
  
\titleformat{\subsection}
  {\normalfont\fontsize{12}{12}\itshape}{\thesubsection}{1em}{}
  
\usepackage[font=small,labelfont=bf]{caption}

\title{\large \textbf{Testing unit root non-stationarity in the presence of missing data in univariate time series of mobile health studies}}

\author[1]{\footnotesize \textbf{Charlotte Fowler}}
\author[2]{\footnotesize \textbf{Xiaoxuan Cai}}
\author[3]{\footnotesize \textbf{Justin T. Baker}}
\author[4]{\footnotesize \textbf{Jukka-Pekka Onnela}}
\author[1]{\footnotesize \textbf{Linda Valeri}}

\affil[1]{Department of Biostatistics, Mailman School of Public Health, Columbia University}

\affil[2]{Department of Statistics, The Ohio State University}

\affil[3]{Institute for Technology in Psychiatry, McLean Hospital, Harvard Medical School}

\affil[4]{Department of Biostatistics, Harvard TH Chan School of Public Health,\\ Harvard University}

\date{\vspace{-4mm} \small October 10, 2022}

\begin{document}

\maketitle

\vspace{-4mm}

\noindent \footnotesize SUMMARY: The use of digital devices to collect data in mobile health (mHealth) studies introduces a novel application of time series methods, with the constraint of potential data missing at random (MAR) or missing not at random (MNAR). In time series analysis, testing for stationarity is an important preliminary step to inform appropriate later analyses. The augmented Dickey-Fuller (ADF) test was developed to test the null hypothesis of unit root non-stationarity, under no missing data. Beyond recommendations under data missing completely at random (MCAR) for complete case analysis or last observation carry forward imputation, researchers have not extended unit root non-stationarity testing to a context with more complex missing data mechanisms. Multiple imputation with chained equations, Kalman smoothing imputation, and linear interpolation have also been proposed for time series data, however such methods impose constraints on the autocorrelation structure, and thus impact unit root testing. We propose maximum likelihood estimation and multiple imputation using state space model approaches to adapt the ADF test to a context with missing data. We further develop sensitivity analysis techniques to examine the impact of MNAR data. We evaluate the performance of existing and proposed methods across different missing mechanisms in extensive simulations and in their application to a multi-year smartphone study of bipolar patients.
\vspace{10mm}


\pagenumbering{gobble}
\clearpage

\pagenumbering{arabic} 

\newpage

\section{Introduction}
\label{s:intro}

Assessing the stationarity of a univariate time series has long been a question of interest to evaluate data assumptions and inform subsequent analyses. However, solutions for testing for stationarity have been largely developed for the field of economics, where the data is typically recorded as a fully observed time series. The increase in the prevalence of time series data across disciplines (e.g., social science, political science, and psychiatry) introduces new challenges for existing time series analysis methods.  

One growing source of time series data is personal digital devices. This technology is increasingly employed to collect information, particularly in the field of health research (Aledavood et. al., 2017; Silva et. al., 2015). It can record participants' real-time exposures and outcomes, with the potential for many observations per second (Torous et. al., 2017; Vaidya et. al., 2013). Mobile health (mHealth) studies for example utilize individuals' smartphones and wearable devices to collect information on the participants daily activities and well-being (Mandel \& Ghosh, 2021; Aledavood et. al., 2017). As smartphones become an essential tool for day-to-day life, mHealth study designs will become ever-more prevalent, and necessitate the development of new statistical methods to handle the particular challenges that mobile health data can introduce. MHealth data encompasses active data which is collected directly from the participant, such as daily survey responses provided by  participants, and passive data which is observed with no action required from the participant, such as GPS records, accelerometer data, and phone and text logs. 
However, these records may suffer from pervasive missingness resulting from technological errors, inactive devices, or participants being unable, unwilling, or uninterested in engaging with the software. 
The Bipolar Longitudinal Study (BLS) is an mHealth cohort study of 74 individuals with Bipolar I or II disorder, schizophrenia, or schizoaffective disorder followed for up to five years using  the Beiwe application (Onnela et. al., 2021, Huang \& Onnela, 2019; Barnett \& Onnela, 2020) on the participant's smartphone. It was configured in the study to record basic information of the participant's anonymized phone calls and text messages, their accelerometer and GPS data, and daily responses to a 31 item survey including questions on mood, sleep, social interaction, physical activity, and other health conditions (Cai et. al., 2022). The study provides researchers with the possibility of identifying the causal effects of at home  interventions such as increased social or physical activity to improve an individual's mood and symptoms. However, the validity of such analysis will be dependent on modeling assumptions, including stationarity. In the Bipolar Longitudinal Study we observe missing data rates for the daily survey varying from 2\% to over 90\%. Missingness is also pervasive in passive data and could result from a variety of factors including glitches, noncompliance, or failure to update software. For example, for one participant in the BLS study we fail to observe any text data for over a four month gap, or 16\% of the follow up their period. Preliminary results in the Bipolar Longitudinal Study also indicate the presence of non-stationarity among several participants' time series data (Cai et. al., 2022). 

In time series analysis, stationarity is defined as constant mean and variance across time (Metcalfe \& Cowpertwait, 2009). Models for stationary processes such as linear mixed models are overwhelmingly used in mHealth (Tewari \& Murphy, 2017; Luckett et. al., 2019; Neto et. al., 2016). However, non-stationarity is likely possible in mHealth studies and it is important to test for it prior to proceeding with traditional longitudinal models. There are many types of non-stationarity such as change points and unstable variance, but the most commonly evaluated is unit root. Unit root non-stationarity is characterized by an autocorrelation of one between the current state and lagged values. A common example of a unit root time series is a random walk. The augmented Dicker Fuller test, or ADF test, is the conventional method used to test for unit root non-stationarity (Metcalfe 
\& Cowpertwait, 2009). However, despite the importance of stationarity assumptions in analyses, unit root testing is often omitted from mHealth studies (Tewari \& Murphy, 2017), likely due to the ADF test requirement for complete data.

There are three classifications for missing data to determine its relative impact on analyses and validity: missing completely at random (MCAR), missing at random (MAR), and missing not at random (MNAR) (Little \& Rubin, 2019).  When data is missing completely at random, the likelihood of a data point not being observed is not influenced by any observed or unobserved data; this is the optimal scenario. In the case of daily surveys from mHealth studies, this could be conceived as a participant failing to respond to survey questions at random a few days a week. Alternatively, when data is MAR, the probability of missingness is only affected by observed variables. An example of missing at random in the mHealth context is loss of interest in study engagement, where missing rates of survey responses increase with time in the study. Lastly, the most challenging mechanism is if data is missing not at random, and the probability of missing data is influenced by unobserved information. For mobile health, this might be represented by an individual not responding to questions on their mood, due to an unobserved depressive state. Because it is hypothesized that mHealth data is often missing at random or missing not random (Goldberg et. al., 2021), time series methods must be adapted to these contexts. 

The existing missing data methods used in time series analysis impose strong assumptions about unit root, and thus may have impacts on validity when used with the ADF test. Basic methods such as mean imputation and last observation carry forward are known to bias estimates and distort variance even when data is MCAR (Little \& Rubin, 2019). Linear and spline interpolation have been shown to produce accurate estimates when used for imputation in covariates (Terry et. al., 1986; Skjelbred and Kong, 2019) but likely bias autocorrelation estimates due to their imposition on the autocorrelation structure. Wijesekara and Liyanage (2020) show under MCAR that Kalman smoothing (Moritz \& Bartz-Beielstein, 2017) achieves high accuracy when imputing values in a univariate time series, however the method has not been evaluated on other missing mechanisms. Commonly used multiple imputation methods developed for non-longitudinal data, such as multiple imputation with chained equations fail to account for the correlation between the lagged and current value when imputing variance (Azur et. al., 2011; Little \& Rubin, 2019), and thus may provide unreliable results. Lastly, complete case analysis which simply drops all unobserved time points and thus disrupts the time sequence may be appropriate when data is MCAR (Shin \& Sarkar, 1996). If the time series has unit root, the autocorrelation will be maintained, while if the autocorrelation is less than one, the autocorrelation will be underestimated. 

The existing research on stationarity testing and autocorrelation estimation with incomplete data is sparse, and relies on strong assumptions about the missingness mechanism. Little and Rubin (2019) propose an EM algorithm for autocorrelation estimation for an AR(1) time series, but condition on stationarity. Shin and Sarkar (1994, 1996) demonstrate that if the time series follows an AR or ARMA structure and data is MCAR, then complete case analysis or last observation carry forward are sufficient methods to be used prior to the ADF test. Park, Shin, et al. (2005) develop a Bayesian test for incomplete data that tests multiple hypotheses of asymmetry and unit root non-stationarity. However there is a need for the investigation of the impact of different missing mechanisms and rates on unit root testing, and a consideration of more sophisticated missing data methods for the ADF test. 

We first propose using imputations from a state space model with multiple imputation (SSMimpute) (Cai et. al., 2022) for unit root testing with the ADF test. While this method is shown to produce non-biased coefficients for covariate estimates with multivariate time series analysis, it has not been tested with a uni-variate time series and estimation of the lagged value coefficient. Under the assumption of a single lag order and normally distributed errors, we additionally propose two likelihood maximization approaches to adapt the ADF test to a context of missing data. The first uses mathematical optimization to maximize the observed data likelihood, and runs the ADF test using the obtained estimates. The second incorporates an EM algorithm to recursively impute missing data and estimate the autocorrelation and variance, and runs the ADF test on the final imputed time series. We additionally introduce sensitivity analyses with a $\delta$ adjustment for the state space model with multiple imputation and maximum likelihood approach using an EM algorithm to consider the ramifications of MNAR data on unit root testing. The remainder of this paper will first review unit root testing  without missing data, then expand these methods to the context of MCAR and MAR data, and lastly introduce sensitivity analyses for MNAR data. We will use simulation and application to BLS mobile health data to review to validity and performance of proposed methods.

\section{Testing unit root without missing data}
\label{s:ts}

Let $t$ represent time point from $t = 0, 1,..., T$. Then assuming lag order of 1, we can define the time series $Y_t$ such that $y_t = \rho y_{t-1} + \epsilon_{t}$, where $E(\epsilon_t) = 0$, $Var(\epsilon_t) = \sigma^2$, and $\epsilon_t$ are independently and identically distributed. We state that time series $y$ has unit root if $\rho = 1$. If this is true, the process is known as a random walk, and as $T\to\infty$, the variance of the time series will diverge to infinity (Metcalfe \& Cowpertwait, 2009). This renders various statistical assumptions invalid, such as those needed to apply standard longitudinal analysis methods including linear mixed models. Notably, preliminary analysis of the Bipolar Longitudinal Study has identified that some participants' response patterns appear to follow a random walk, demonstrating the importance of testing for unit root in mHealth data (Cai et. al., 2022). The Dickey Fuller test was developed to test the null hypothesis: $H_0: \rho = 1$ against the alternative that $H_1: \rho <1$ (Dickey \& Fuller, 1979). The test employs ordinary least square methods regressing $Y_t$ by $Y_{t-1}$to estimate $\hat\rho$ and $\hat\sigma^2$ such that test statistic is defined as: 

\begin{equation}
    DF_{\hat\rho} =  
    \frac{\hat\rho-1}{SE(\hat\rho)} = 
    \left(\frac{\sum{y_t y_{t-1}}}{\sum y_{t-1}^2}-1\right)\Bigg/\sqrt{\frac{T\hat\sigma^2}{T\sum{y_{t-1}^2-(\sum y_{t-1})^2}}}
    .
\end{equation}

Under the null hypothesis, the distribution of the test statistic has no closed form (Dickey \& Fuller, 1979), but is named the Dickey Fuller distribution. For testing purposes, p-values and rejection regions are generated through interpolation, using a table of distribution quantiles obtained via simulation (Dickey \& Fuller, 1979). Assuming $\epsilon_t$ are distributed normally, we note that $y_t|y_{t-1}\sim N(\rho y_{t-1}, \sigma^2)$.  We can thus solve for the joint log-likelihood of $\rho, \sigma^2$ conditional on $y_1, ..., y_T$ as: 

\begin{equation}
    \ell(\rho, \sigma^2|y_1,...,y_T) = 
    \sum_{t = 1}^{T} \log f(y_{t}|y_{t-1}) =  
    -T/2\log(2\pi\sigma^2) - \frac{1}{2\sigma^2}\sum_{t = 1}^T(y_t-\rho y_{t-1})^2.
\end{equation}


 

\section{Testing unit root in the presence of MCAR or MAR missing data}
\label{s:mis}

For $t = 1,...,T$, let $r_t$ serve as an indicator of whether $y_t$ is observed, and  $p_t$ represent the probability of observing $y_t$, i.e. $p_t = P(r_t =1)$. If data is missing completely at random, we assume that $p_t$ is independent of $Y_t, t$, and any other observed or unobserved variables, for $t = 1,...,T$. When the time series is missing at random, $p_t$ is dependent only on observed variables. In the case of a univariate time series, we will assume that the probability of observing a time point is dependent on only observed covariates $X$ including time $t$ and previously observed non-missing $y_t$, such that for some function $g$, $p_t = g(X)$. We let $k = t_1,...,t_n$ represent the index of times with non-missing observations, such that $y_{t_1}, y_{t_2}, ..., y_{t_n}$ represent all observed time series values, with $n$ total non-missing observations, and $n\leq T$. Note that as seen in (Bertin et. al., 2011), we can solve that the distribution of $y_{t_k}$ conditional on the previously observed point $y_{t_{k-1}}$ is normal, such that 

\begin{equation}
    y_{t_k}|y_{t_k-1}\sim N(\rho^{t_k-t_{k-1}}y_{t_k-1}, \sigma^2V_k(\rho)) \textnormal{, with }
    V_k(\rho) = \sum_{j = 1}^{t_k - {t_{k-1}}} \rho^{2(j-1)}. 
\end{equation}

We can now adapt the complete time series log-likelihood from (2) for observed data: 
\begin{align*}
   \ell(\rho, \sigma^2|y_{t_1},...,y_{t_n}) =& 
    \sum_{t = 1}^T \left(r_t\log p_t + (1-r_t)\log(1-p_t)\right) + \sum_{k = 1}^n\log f(y_{t_k}|y_{t_{k-1}}) \\
    =& \sum_{t = 1}^T \left(r_t\log p_t +  (1-r_t)\log(1-p_t)\right)-n/2\log(2\pi\sigma^2) \numberthis \label{eqn}\\ 
    &\quad - \sum_{k = 1}^n\left(\frac{1}{2}\log V_k(\rho)-\frac{y_{t_k}- \rho^{(t_k - {t_{k-1}}y_{t_{k-1}})^2}}{2\sigma^2V_k(\rho)}\right). 
\end{align*}
    
Note that when $p_t$ is independent of $\rho$ and $\sigma$, the likelihood can be maximized with respect to these parameters by disregarding the first term of the log-likelihood. This condition holds when data is MCAR or MAR.


\subsection{Maximum Likelihood Estimation with Numeric Optimization}

Using likelihood from (4), where we assume $\epsilon_t$ is distributed normally, even if we assume data is MCAR or MAR and ignore the first term, we are unable to solve for an algebraic maximum. Instead, we suggest using numerical optimization, specifically the Nelder-Mead method to maximize the function with respect to $\sigma^2$ and $\rho$. We calculate initial estimates of $\rho$ and $\sigma$ using only time points where $y_t$ and $y_{t-1}$ are observed, such that 

\begin{equation}
\hat\rho_{(0)} =  \frac{\sum_{t = 1}^T r_t r_{t-1}y_t y_{t-1}}{\sum_{t = 1}^T r_t r_{t-1} (y_{t-1})^2} \textnormal{ \quad and \quad } \hat\sigma^2_{(0)}  =  \frac{\sum_{t = 1}^T r_t r_{t-1} (y_t - y_{t-1})^2}{\sum_{t = 1}^T r_t r_{t-1}}.
\end{equation}

To implement the ADF test using estimates generated from numerical maximization ($\hat{\rho}, \hat{\sigma^2}$), we calculate a conservative new test statistic as: 

\begin{equation}
    DF_{\hat\rho, c} = \frac{\hat\rho - 1}{SE(\hat\rho)} = (\hat\rho - 1 ) \Bigg/\sqrt{\frac{n\hat{\sigma}^2}{n\sum_{k = 1}^{n-1}y_{t_k}-\left(\sum_{k = 1}^{n-1}y_{t_k}\right)^2}}.
\end{equation}

We will refer to this method as MLEN (maximum likelihood estimation by numerical optimization, conservative). This statistic fails to leverage on the longer observation period, and treats the follow up time as strictly the number of non-missing time points. For time series with severe missing rates, or shorter follow up, the conservative nature of this test statistic becomes a limitation for its power. To address this issue, we additionally consider a scaled version of the above statistic, where we leverage on the unobserved time points, by calculating $DF_{\hat\rho, s} = \frac{T}{n} DF_{\hat\rho, c}$. We will refer to this method as  MLENS (maximum likelihood estimation by numerical optimization, scaled). 

When assumptions of normality and lag order of one are satisfied, the algorithm benefits from exact model specification, allowing for accurate results with computational ease.


\subsection{Maximum Likelihood Estimation with Expectation Maximization algorithm}

We additionally consider a maximum likelihood estimation approach using an iterative expectation maximization algorithm (MLEEM). We calculate initial estimates for the parameters as in (5). The algorithm then recursively imputes missing observations, and updates parameters until convergence. In iteration $j$, we impute missing observations chronologically such that for unobserved $y_t$,  $\hat{y}_t^{(j)} = \hat\rho_{(j-1)} \hat{y}_{t-1}^{(j)}$. In the M step, we maximize the likelihood using the imputed time series, which is equivalent to ordinary least squares estimation when regressing $Y_t$ by $Y_{t-1}$, and update estimates $\hat\rho_{(j)}$ and $\hat\sigma^2_{(j)}$. 

 Similarly to the numerical optimization method, the EM approach for maximum likelihood estimation benefits from a exact model specification, and computational efficiency when assumptions hold. It additionally allows for direct calculation of the ADF test statistic from the fully imputed time series, avoiding issues of reduced power encountered by numerical optimization without imputation. 


\subsection{State Space Model with  Multiple Imputation}

We lastly propose a more flexible imputation method to be used prior to unit root testing. The state space model with multiple imputation (SSMimpute) does not assume normality, nor a single order lag structure (Cai et. al., 2022). Assuming $q$ lag order, the state space model is fit by regressing $Y_t$ by its lagged values, $Y_{t-q},..., Y_{t-1}$. We generate initial estimates for the missing values of $Y_t$ using Kalman filtering, and impute these values in the lagged regressors, $Y_{t-q},..., Y_{t-1}$, but not $Y_t$. The algorithm then recursively until convergence estimates the coefficients of the regressors and the variance through maximizing the likelihood and  calculating new imputations from the posterior to update missing values of $Y_{t-q},..., Y_{t-1}$. Note that in the case of a single lag order, the maximization step will be estimating $\rho$ and $\sigma^2$ as defined previously. When convergence is achieved for the likelihood and coefficient estimation, multiple imputations are drawn from the last posterior distribution. These imputations are used to conduct the ADF test.  Because the Dickey Fuller test statistic's distribution has no closed form, we pool test results across imputations by calculating the median test statistic, as has been employed in other contexts (Eekhout et. al., 2017; van de Wiel et. al., 2009). Specifically, the SSMimpute method (Cai et. al., 2022) for unit root testing follows the following procedure: 

\begin{enumerate}
    \item Initialization: Generate initial imputations for missing values $\hat Y_t^{(0)}$, and substitute imputations into the corresponding missing lagged values to eliminate missingness in the regressors.
    
    \item Maximization: In the $k$th iteration, apply the state space model to outcome with missing values $Y_t$ and the explanatory imputed lagged $\hat Y_{t-q}^{(k-1)},..., \hat Y_{t-1}^{(k-1)}$ to obtain maximum likelihood estimates for the coefficients of regressors. 
    
    \item Substitution: Calculate imputations of $Y_t^{(k)}$ from the updated posterior distribution and substitute these imputations into the corresponding lagged variables $\hat Y_{t-q}^{(k-1)},..., \hat Y_{t-1}^{(k-1)}$.
    
    \item Check convergence: Repeat steps 2 and 3 until convergence is reached for likelihood and coefficient estimation.
    
    \item Multiple imputation: Once convergence is achieved in iteration $K$, obtain $M$ random draws of coefficient estimates from the posterior distribution reached in iteration $K$. From each set of coefficient estimates, calculate imputations  $\hat Y_t^{(K,m)}$. 
    
    \item Unit root testing: Apply ADF test to each imputation  $\hat Y_t^{(K,m)}$ for $m = 1, ..., M$ and obtain the resulting test statistic. Pool across imputation results by calculating the median test statistic and its corresponding p-value. 
\end{enumerate}

The SSMimpute approach for unit root testing relaxes assumptions by not relying on normality and single lag order. Although this makes it slightly more computationally intensive, by not calculating multiple imputations within each iteration, it eases the computational burden without compromising on estimate accuracy (Cai et. al., 2022). Additionally Cai et. al. (2022) demonstrated the method has low bias across a range of data generation scenarios. 

Full properties of the four proposed methods for testing unit root non-stationarity with missing data can be seen in Table 1. 

\section{MNAR sensitivity analyses}
\label{s:nmar}
Data is classified as missing not at random, or MNAR when the probability of not observing a time point is dependent on unobserved information. This is the most severe missing data classification, and yet it cannot be tested for due to its nature. In the case of a univariate time series, we assume that data MNAR would signify that the probability of missing a time point is dependent on the value of that individual time point. In the psychiatric mobile health context, this missingness seems plausible, with participants being less likely to report a negative mood when in a depressive state. Because we are unable to observe or test for the influence of missing not at random, sensitivity analyses are commonly used to assess the impact of MNAR data, through incorporating a range of values of  $\delta$ which represent the hypothesized difference between observed and unobserved time points. We propose methods to conduct such a sensitivity analysis for unit root testing for the maximum likelihood estimation with expectation maximization and state space model with multiple imputation methods. 

The MLEEM method's sensitivity analysis within each iteration, recursively across time $t$ adds a term to each imputation for missing value $y_t$ such that $\hat{y}_t^{(j)} = \hat\rho_{(j-1)} \hat{y}_{t-1}^{(j)} \pm \delta_t$. Thus $\delta_t$ represents the hypothesized missing mechanism, or specifically the expected difference in $y_t$ given $r_t = 1$ and $y_t$ given $r_t = 0$. As our eventual goal is testing for unit root, we wish avoid creating any major jumps in the time series when imputing, to not disrupt autocorrelation estimates. To do so, we assume that for a missing gap from time points $y_{u}$ to $y_{v}$, the time series will add $\delta_t$ to imputations for the first half of time points $t = (u, u+1, ..., u + \lfloor\frac{v-u}{2}\rfloor)$, and subtract $\delta_t^{(t)}$ for time points $t = (u + \lfloor\frac{v-u}{2}\rfloor +1 ,u + \lfloor\frac{v-u}{2}\rfloor +2, ..., v)$. This will essentially create a rise and fall peak effect across each missing gap. 

We focus on the context where $\delta_t$ is constant across $t$, however we additionally consider the case from Example 15.4 in Little and Rubin (2019) where it is hypothesized that all missing values fall between $(\lambda, \infty)$. Note that in this scenario, for a sensitivity analysis $\lambda$ would be chosen as a range of plausible values which would subsequently inform $\delta_t$. Under this hypothesis, conditioning on the normality assumption we can further inform our  $\delta$. We let $\phi$ and $\Phi$ represent the standard normal density and cumulative distribution functions respectively, and obtain the following adjustment: 

$$\delta_t^{(j)} = \frac{\sigma^{(j)} \phi\left(\frac{\lambda-\hat\rho_{(j-1)} \hat{y}_{t-1}^{(j)}}{\sigma^{(j)}}\right)}{1- \Phi\left(\frac{\lambda-\hat\rho_{(j-1)} \hat{y}_{t-1}^{(j)}}{\sigma^{(j)}}\right)}.$$

For the SSMimpute approach, following initialization, we incorporate $\delta$ within the E-step of each iteration of the algorithm. After  imputation of missing values in lagged regressors, we will add a $\delta$ informed term to ensure all imputations are shifted to represent the hypothesized MNAR effects. Thus if the state space model in iteration $j$ imputes for lagged regressor $Y_{t-q}$ at time point $t$ a value $\tilde{y}_{t-q}^{(j)}$, we calculate the final imputation in this iteration to be $\hat{y}_{t-q}^{(j)} = \tilde{y}_{t-q}^{(j)} + m_{t-q}\delta_{t-q}$. Similarly to the MLEEM sensitivity analysis, we wish to create a rise and fall of imputed values, to avoid a large jump in the imputed time series. Thus for a missing gap in between time points $u$ and $v$, we let $m_t = (t - u +1)$ for $t = (u, u+1, ..., u + \lfloor\frac{v-u}{2}\rfloor)$, and $m_t = (v-u -t)$ for $t = (u + \lfloor\frac{v-u}{2}\rfloor +1 ,u + \lfloor\frac{v-u}{2}\rfloor +2, ..., v)$. We will refer to this adjustment as peak $\delta$. We  additionally consider the case where a value missing has a stagnant effect, and let $m_t = 1$ across all time points ($\delta_s$). As by the time the algorithm converges the $\delta$-adjustment should be incorporated into the posterior, we do not additionally impose the adjustment to the multiple imputations sampled from said posterior.

\begin{table}[]
\caption{Properties of four proposed methods for unit root testing with missing data.}
\footnotesize
    \centering
    \setlength{\tabcolsep}{4pt}
    \setlength{\extrarowheight}{8pt}

    \begin{tabular}{l l l l l}
    \hline
        & 
      \parbox[t]{3.5cm}{\bf{MLE with\\Expectation\\Maximization \\(MLEEM)}}  & 
      \parbox[t]{3.5cm}{\bf{MLE with\\Numeric\\ Optimization\\ (MLEN)}}  & 
      \parbox[t]{3.5cm}{\bf{MLE with\\Numeric\\ Optimization,\\ scaled\\(MLENS)}}  & 
      \parbox[t]{3cm}{\bf{State Space\\Model with\\ Multiple\\Imputation\\(SSMimpute)}}  \\ \hline
      \bf{Assumptions} & 
      \parbox[t]{3.5cm}{Errors are independently and identically normally  distributed;\\Lag order of one}  & 
      \parbox[t]{3.5cm}{Errors are independently and identically normally  distributed;\\Lag order of one}  & 
      \parbox[t]{3.5cm}{Errors are independently and identically normally  distributed;\\Lag order of one} & 
      \parbox[t]{3cm}{Continuous $y_t$;\\Finite lag order}\\ & \\  \hline
      \bf{Advantages} &
      \parbox[t]{3.5cm}{Computational \\efficiency;\\Able to incorporate $\delta$ for sensitivity analyses}  & 
      \parbox[t]{3.5cm}{Computational \\efficiency; \\Very low type I error}  & 
      \parbox[t]{3.5cm}{Computational \\efficiency} &
      \parbox[t]{3cm}{Flexible modeling assumptions;\\Able to incorporate  $\delta$ for sensitivity analyses}\\ & \\  \hline
      \bf{Limitations} &
      \parbox[t]{3.5cm}{Strict assumptions}  & 
      \parbox[t]{3.5cm}{Strict assumptions;\\Relatively low power}  & 
      \parbox[t]{3.5cm}{Strict assumptions}  & 
      \parbox[t]{3cm}{Relatively\\computationally intensive}  \\ & \\ \hline
    \end{tabular}
    
\end{table}

\section{Simulation Settings}
\label{s:sim design}

To assess the performance of proposed methods against existing missing data approaches for univariate time series, we conduct a simulation study, divided in two parts. The first applies methods under the hypothesis that data is MCAR or MAR, and the later employs the proposed sensitivity analyses, under the assumption data is MNAR. 

\subsection{Main Simulation}

We conduct 2000 simulations generating time series with $T = 500$ and a single lagged order such that $y_{t} = \rho y_{t-1} + \epsilon_t$ and $\epsilon_t\sim N(0, 1)$. For each simulation, we consider $\rho = .5, .9, .95, 1$. Within each simulation for the four time series we consider MCAR, MAR, MNAR missing mechanisms, with missing rates of 30, 50, and 70 percent. For MCAR data we determine missingness by a random sample of $T$ from a binomial distribution with fixed $p = 0.3, 0.5, 0.7$. For MAR, we specify $p = c \times t$, with $c$ set such that on average the desired missing rate was achieved, and the probability of missing increases as $t$ increases. Lastly, to simulate MNAR, we first consider an extreme deterministic scenario, where values of $y_t$ above the 30, 50, or 70th quantile are all classified as missing (MNAR-D). We additionally for this simulation consider data such that probability of missing, $p$ is calculated as an interaction between $t^2$ and $y_t$ (MNAR-T). Specifically, we calculate $p= e^{c t^2 y_t}/ (1+ e^{c t^2 y_t })$, with $c$ defined such that on average the desired missing rate is achieved. For each resulting time series, we apply mean imputation (M), last observation carry forward (LOCF), linear and spline interpolation (IntL, IntS)and Kalman smoothing imputation (K) from the imputeTS package in R (Moritz \& Bartz-Beielstein, 2017), and complete case analysis (CC) including only observed time points. For each of these single imputations and complete case analysis we apply the Dickey Fuller test. We additionally consider multiple imputation with chained equations (MICE) with covariates $y_{t-1}, t$ and $m = 5$ from the mice R package (van Buuren \& Groothuis-Oudshoorn, 2011), and obtain a pooled result for the Dickey Fuller test by taking the median test statistic. We compare these methods with results from MLEN, MLENS, MLEEM, and SSMimpute with $m = 5$, and a single lag order. In addition to extracting the test statistic and p-value for each simulation, we calculate the estimated autocorrelation, $\hat{\rho}$. To evaluate the methods, we examine autocorrelation, test statistic, and p-value distributions across $\rho$, and the power across different values of $\rho$ and type 1 error.

\subsection{Sensitivity Analysis Simulation} 

To evaluate proposed sensitivity analyses methods, we ran 1000 simulations with T = 500, for $\rho = .5, .9, .95, 1$. For the simulation we consider 4 different MNAR mechanisms. First, we consider MNAR-D and MNAR-T, as described above. We additionally consider, MNAR-P, which uses a probabilitic framework such that the time series observations were divided into three quantiles with 40\%, 10\%, and 0\% missing rates, in descending order of quantile. The fourth, MNAR-H, a hybrid between probabilistic and deterministic missing, added an additional quantile to MNAR-P in which all data was missing. As in the main simulation, we consider 30, 50, and 70 percent missing rates, and set quantile cutoffs accordingly. For MLEEM, we consider $\delta_t = 0.05, 0.1, 0.2, 0.3$ with a fixed $\delta$ across time. For SSMimpute, we consider a peak $\delta_t = 0.05, 0.1, 0.2, 0.3$. We remove the stagnant $\delta$ adjustment for SSMimpute due to convergence issues.

\section{Simulation Results}
\label{s:sim res}

The percent of simulations which rejected the null of unit root in the main simulation are shown in Table 2 by missing rate, method, missing mechanism (MCAR, MAR, MNAR-D), and $\rho$. For $\rho = 1$, the value represents type I error, while for $\rho<1$, the value is equivalent to power. The entry with rate of 0 for each $\rho$ under complete case (CC) represents the performance of the ADF test with no missing data. We omit the case where $\rho = 0.5$, as here all methods have power greater than 0.95. Additionally, MICE and mean imputation are omitted, but across all simulations both methods had a type I error greater than 0.9, demonstrating their ineffectiveness for unit root testing. The MLEN method is indeed very conservative, with type I error of 0 for MCAR and MAR data, and low power. When data is MCAR or MAR, MLEEM, MLENS, and SSMimpute have power over 79\% for a $\rho$ of 0.95. These methods additionally demonstrate type I error less that 0.05 with a missing rate of 30\% when data is MCAR or MAR. Complete case analysis also has comparable type I error, but power suffers with higher rates of missing data, as expected. When data is MAR with 70\% missing, the complete case power is merely 32\%, compared to 96\% and 85\% for MLENS and SSMimpute, respectively. We observe that LOCF exhibits both inflated type I error and reduced power, while linear and spline (excluded from table) interpolation and Kalman smoothing imputation both bias towards non-stationarity and have low power. For a deterministic missing not at random, or MNAR-D, time series, all methods perform poorly as expected, given they are misspecified. Regardless of method, even with only 30\% missing not at random, type I error is greater than 8\%, meaning an increase chance of rejecting the null of non-stationarity when the time series is truly non-stationary. With SSMimpute this effect is especially severe, with a type I error of 30\%. For time series with probability of missing as an interaction of value and time, MNAR-T, and moderate missing rates, methods no longer exhibit an inflated type I error, but power does suffer, particularly for complete case analysis.

\begin{table}[]
\renewcommand{\arraystretch}{0.9}
\footnotesize
\caption{\footnotesize Simulated power (when $\rho<1)$ and type I error (when $\rho = 1$) by missing mechanism, $\rho$, missing rate, and method. We denote complete case analysis by CC, SSMimpute by SSM, last observation carry forward by LOCF, linear interpolation by IntL, and  Kalman smoothing imputation by K.}
\begin{tabular}{@{}lllllllllll@{}}
\toprule
 & $\rho$ & Rate & CC & MLEEM & MLEN & MLENS & SSM & LOCF & IntL & K \\ \midrule
MCAR & 1 & 0 & 0.05 &  &  &  &  &  &  &  \\
 &  & 0.3 & 0.05 & 0.05 & 0 & 0.01 & 0.05 & 0.23 & 0.01 & 0.05 \\
 &  & 0.5 & 0.06 & 0.05 & 0 & 0.01 & 0.06 & 0.21 & 0.01 & 0.06 \\
 &  & 0.7 & 0.08 & 0.05 & 0 & 0.02 & 0.08 & 0.21 & 0.01 & 0.08 \\
 & 0.95 & 0 & 0.83 &  &  &  &  &  &  &  \\
 &  & 0.3 & 0.81 & 0.82 & 0.65 & 0.92 & 0.83 & 0.81 & 0.36 & 0.38 \\
 &  & 0.5 & 0.78 & 0.81 & 0.42 & 0.97 & 0.83 & 0.77 & 0.08 & 0.10 \\
 &  & 0.7 & 0.72 & 0.80 & 0.04 & 0.98 & 0.85 & 0.67 & 0 & 0.08 \\
 & 0.90 & 0 & 1 &  &  &  &  &  &  &  \\
 &  & 0.3 & 1 & 1 & 1 & 1 & 1 & 1 & 0.99 & 0.99 \\
 &  & 0.5 & 1 & 1 & 0.99 & 1 & 1 & 1 & 0.80 & 0.84 \\
 &  & 0.7 & 1 & 1 & 0.63 & 1 & 1 & 1 & 0.10 & 0.26 \\ \midrule
MAR & 1 & 0 & 0.05 &  &  &  &  &  &  &  \\
 &  & 0.3 & 0.05 & 0.05 & 0 & 0.01 & 0.05 & 0.17 & 0.02 & 0.02 \\
 &  & 0.5 & 0.05 & 0.06 & 0 & 0.05 & 0.07 & 0.13 & 0.01 & 0.02 \\
 &  & 0.7 & 0.04 & 0.11 & 0 & 0.10 & 0.33 & 0.08 & 0.05 & 0.07 \\
 & 0.95 & 0 & 0.83 &  &  &  &  &  &  &  \\
 &  & 0.3 & 0.83 & 0.82 & 0.13 & 0.95 & 0.83 & 0.79 & 0.36 & 0.39 \\
 &  & 0.5 & 0.72 & 0.82 & 0.76 & 1 & 0.84 & 0.70 & 0.12 & 0.2 \\
 &  & 0.7 & 0.32 & 0.79 & 0.76 & 0.96 & 0.85 & 0.70 & 0.24 & 0.3 \\
 & 0.90 & 0 & 1 &  &  &  &  &  &  &  \\
 &  & 0.3 & 1 & 1 & 1 & 1 & 1 & 1 & 0.99 & 0.99 \\
 &  & 0.5 & 1 & 1 & 1 & 1 & 1 & 1 & 0.84 & 0.90 \\
 &  & 0.7 & 0.83 & 1 & 0.7 & 1 & 1 & 0.99 & 0.81 & 0.87 \\ \midrule
MNAR-D & 1 & 0 & 0.05 &  &  &  &  &  &  &  \\
 &  & 0.3 & 0.12 & 0.13 & 0.08 & 0.17 & 0.30 & 0.17 & 0.10 & 0.12 \\
 &  & 0.5 & 0.16 & 0.22 & 0.06 & 0.20 & 0.54 & 0.20 & 0.19 & 0.19 \\
 &  & 0.7 & 0.18 & 0.35 & 0.03 & 0.19 & 0.82 & 0.39 & 0.39 & 0.31 \\
 & 0.95 & 0 & 0.83 &  &  &  &  &  &  &  \\
 &  & 0.3 & 0.94 & 0.96 & 0.88 & 0.96 & 0.98 & 0.94 & 0.92 & 0.93 \\
 &  & 0.5 & 0.91 & 0.96 & 0.44 & 0.81 & 0.99 & 0.97 & 0.96 & 0.95 \\
 &  & 0.7 & 0.77 & 0.91 & 0.03 & 0.33 & 0.99 & 0.98 & 0.97 & 0.78 \\
 & 0.90 & 0 & 1 &  &  &  &  &  &  &  \\
 &  & 0.3 & 1 & 1 & 0.99 & 1 & 1 & 1 & 1 & 1 \\
 &  & 0.5 & 1 & 1 & 0.60 & 0.96 & 1 & 1 & 1 & 1 \\
 &  & 0.7 & 0.98 & 1 & 0.01 & 0.40 & 1 & 1 & 1 & 0.99 \\ \midrule
MNAR-T & 1 & 0 & 0.05 &  &  &  &  &  &  &  \\
&  & 0.3 & 0.05 & 0.05 & 0 & 0.01 & 0.06 & 0.16 & 0.02 & 0.02 \\
&  & 0.5 & 0.06 & 0.07 & 0 & 0.05 & 0.13 & 0.13 & 0.03 & 0.04 \\
&  & 0.7 & 0.05 & 0.19 & 0 & 0.10 & 0.55 & 0.13 & 0.13 & 0.16 \\
& 0.95 & 0 & 0.83 &  &  &  &  &  &  &  \\
&  & 0.3 & 0.80 & 0.83 & 0.84 & 0.97 & 0.84 & 0.80 & 0.39 & 0.44 \\
&  & 0.5 & 0.55 & 0.81 & 0.60 & 0.97 & 0.82 & 0.71 & 0.35 & 0.40 \\
&  & 0.7 & 0.18 & 0.80 & 0.08 & 0.87 & 0.88 & 0.72 & 0.47 & 0.51 \\
& 0.90 & 0 & 1 &  &  &  &  &  &  &  \\
&  & 0.3 & 1 & 1 & 1 & 1 & 1 & 1 & 0.99 & 1 \\
&  & 0.5 & 0.99 & 1 & 0.99 & 1 & 1 & 0.99 & 0.95 & 0.97 \\
&  & 0.7 & 0.62 & 1 & 0.48 & 1 & 1 & 0.99 & 0.93 & 0.95 \\
\bottomrule
\end{tabular}
\end{table}

The box-plots of p-values by method and missing mechanism with 50\% missing in Figure 1 further demonstrate simulation results. High power is seen by box-plots for $\rho<1$ mostly below the $\alpha = 0.05$  cutoff, and low type I error is visualized by little of the $\rho = 1$ distribution of p-values below the $\alpha = 0.05$ line. Across all methods the MNAR-T mechanism does not suffer from the high type I error seen for MNAR-D, but does have lower power compared to MCAR and MAR.  In Figure 2, the $\hat\rho$ estimates by method are plotted, demonstrating the biases by method and missing mechanism. The four proposed methods, SSMimpute, MLEEM, MLEN, and MLENS all have low bias when data is MCAR or MAR. As seen with p-values, mean and MICE imputation underestimate $\rho$. LOCF, linear interpolation, and Kalman smoothing imputation  bias results towards non-stationarity, as is seen by their estimates above the true value of $\rho$, when $\rho < 1$.  As expected, complete case analysis underestimates $\rho$ when $\rho<1$, and is accurate when $\rho=1$. In Web Appendix A, we include graphs for the p-values, $\hat\rho$ estimates, and test statistics with 30, 50, and 70 percent missing across methods, $\rho$, and missing mechanism.

\begin{figure}
\centerline{%
\includegraphics[width = 170mm]{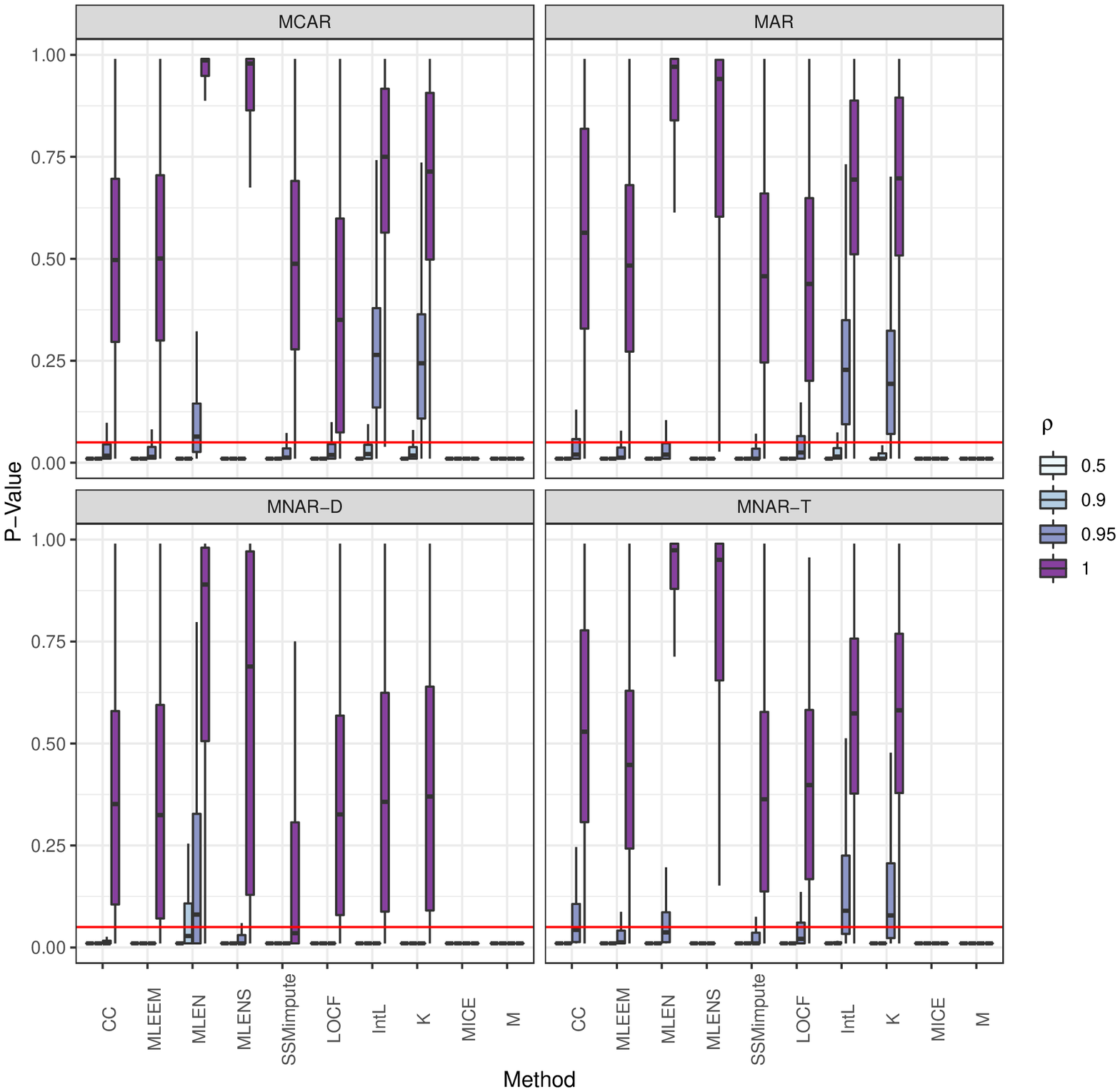}}
\caption{ P-value box plots by missing mechanism, $\rho$, and method for time series with 50\% missing. The red horizontal line indicates cut-off values of $\alpha = 0.05$. We denote complete case analysis by CC, last observation carry forward by LOCF, linear interpolation by IntL,  Kalman smoothing imputation by K, and mean imputation by M.} 
\label{Figure 1}
\end{figure}

\begin{figure}
\centerline{%
\includegraphics[width = 170mm]{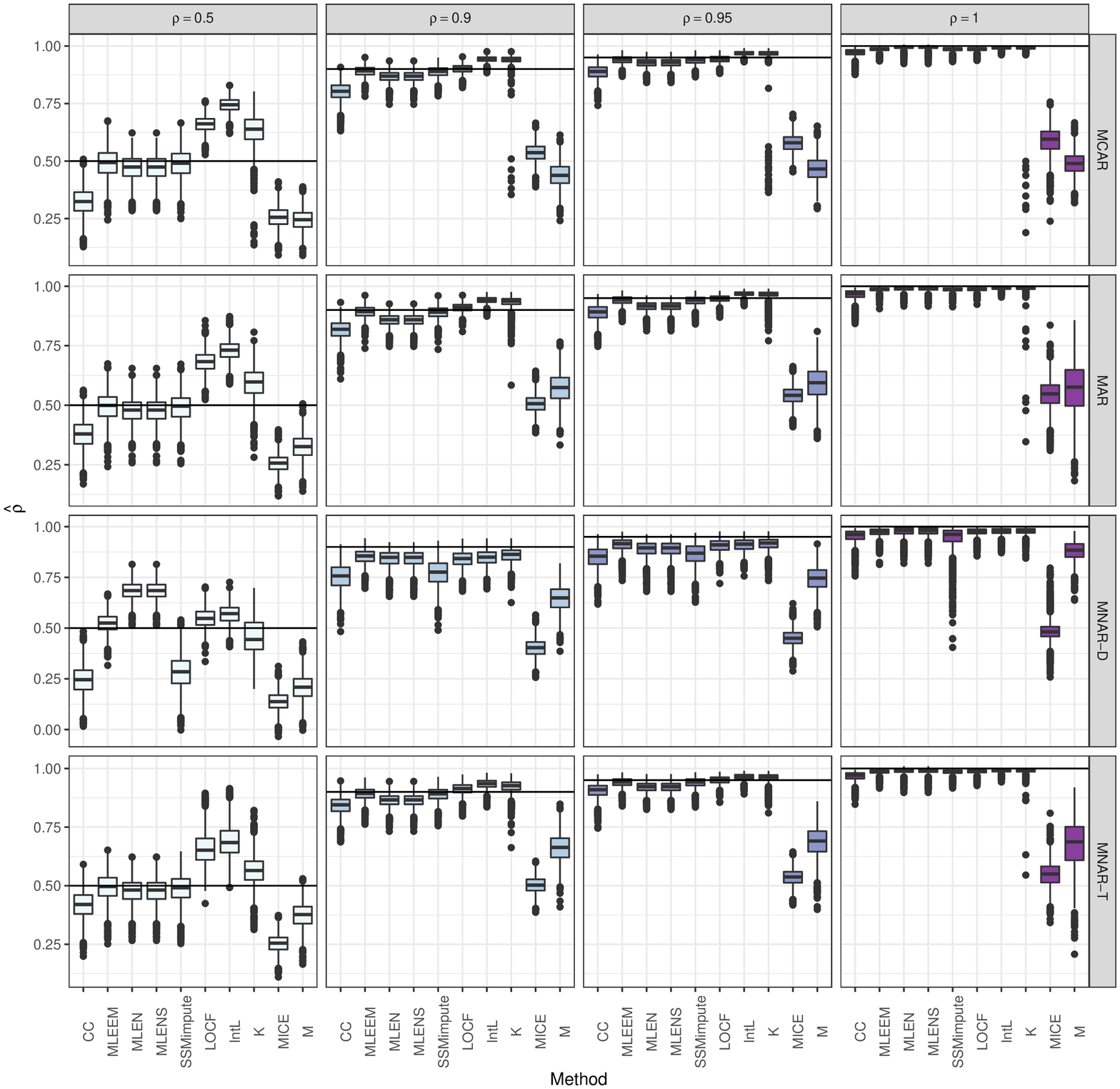}}
\caption{Autocorrelation estimates for time series with 50\% missing by missing mechanism, true $\rho$, and method. Black horizontal lines indicate the true value of $\rho$.}
\label{Figure 2}
\end{figure}

In the sensitivity analysis simulation, we see that across different types of MNAR data the proposed $\delta$ adjustments are able to shift results from the ADF test. Specifically, we demonstrate that with positive $\delta$ values, we are able to correct the inflated type I error exhibited by MNAR-D data in the general simulation. In Figure 3, p-values for the SSMimpute and MLEEM methods are visualized across a range of $\delta$ values for MNAR-D time series. Increases in $\delta$ generate a increase in p-value for both methods, representing a reduced likelihood of rejecting the null hypothesis and thus a shift towards non-stationarity. Simulation results additionally demonstrate that the SSMimpute method appears more sensitive to changes in $\delta$ compared to MLEEM. This was expected, as it is a more flexible model and thus more able to adapt to reflect changes in the imputation structure. The results for hybrid and probibalistic missing not at random are similar to those seen with deterministic missing not at random.  For MNAR-T, we did not see the same pattern of inflated type I error in prior to sensitivity analyses, and the proposed $\delta$ adjustments tested fail to incorporate the impact of $t$ on the probability of missing. Despite these limitations, the increasing $\delta$ adjustments do provide an increasing range of p-value and autocorrelation estimates. This is especially pronounced when applied to time series with higher rates of missing data. Full visualizations across tested MNAR mechanisms are shown in Web Appendix B, with autocorrelation and p-value graphs by missing mechanism, missing rate, and $\rho$. 

\begin{figure}
\centerline{%
\includegraphics[width = 170mm]{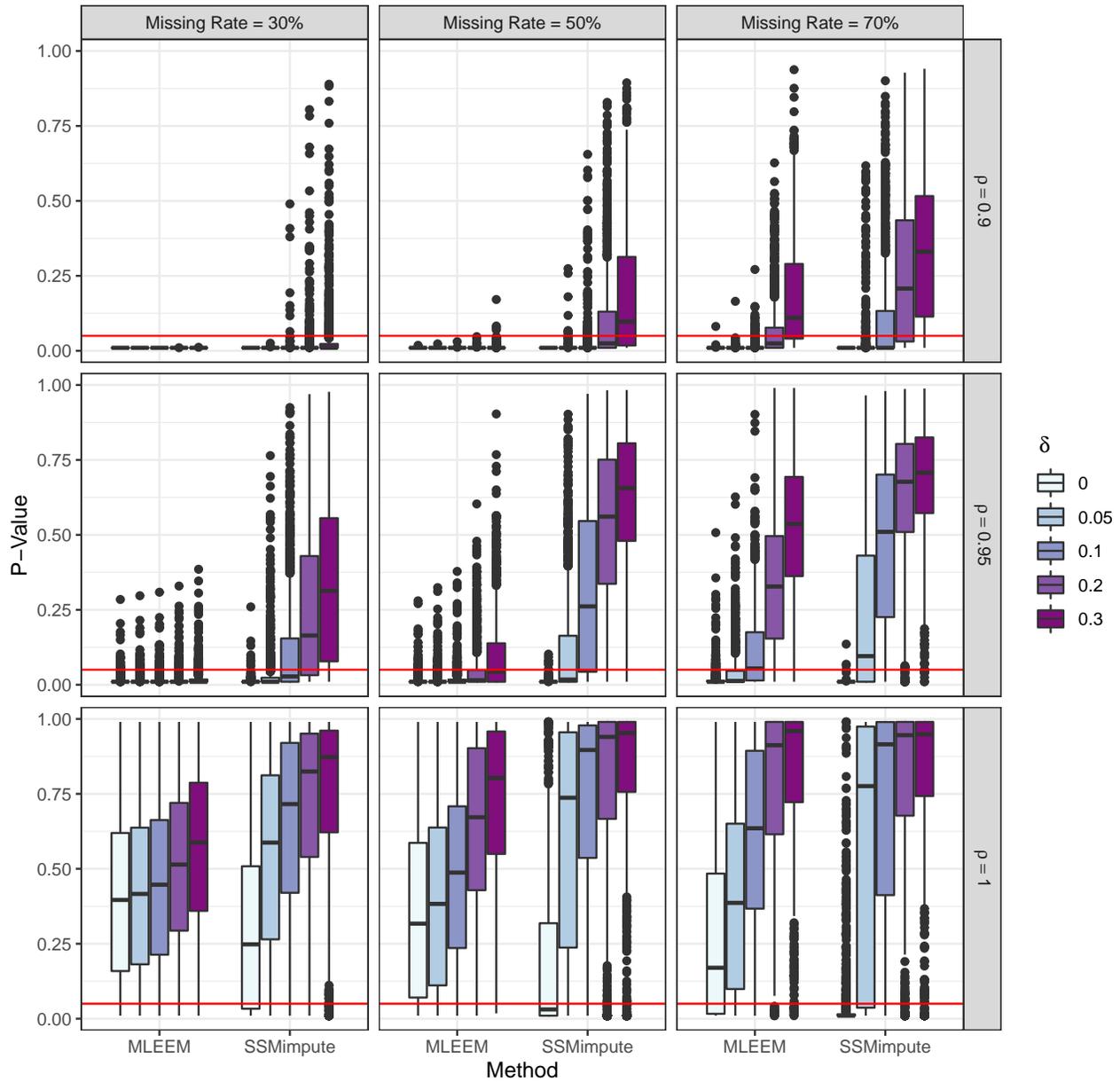}}
\caption{P-value box plots for sensitivity analysis by missing rate, true $\rho$, method, and $\delta$ adjustment. The red horizontal line indicates cut-off values of $\alpha = 0.05$.}
\label{Figure 3}
\end{figure}

Overall when the time series is MCAR or MAR, the simulation demonstrates the high power and acceptable type I error of proposed methods, and the limitations of existing methods when testing for unit root non-stationarity. We additionally observe inflated type I error across all methods when data is MNAR. The impact of time points missing not at random can be assessed using a sensitivity analysis with the MLEEM or SSMimpute methods by specifying a range of $\delta$ adjustments.

\section{Application to Mobile Health Data}
\label{s: application}

We applied the proposed methods for unit root testing in univariate time series with missing data to mobile health data collected from the Bipolar Longitudinal Study (BLS).  BLS is an ongoing longitudinal cohort study of 74 participants with Bipolar I or II disorder, schizophrenia, or schizoaffective disorder recruited from the Psychotic Disorders Division at McLean Hospital in Belmont, MA. The Beiwe mobile application developed by Jukka-Pekka Onnela's lab (Onnela et. al., 2021, Huang \& Onnela, 2019; Barnett \& Onnela, 2020) is employed to passively collect data on physical activity, GPS locations, and telecommunications of texts and calls. The application additionally was configured to prompt users to daily respond to a 5-minute survey at 5:00 PM on their moods, sleep, social activity, and psychotic symptoms. To apply proposed methods, we focus on negative and positive mood scores which are built as aggregates of survey responses to questions. Negative mood score is generated from questions relating to fear, anxiety, embarrassment, hostility, stress, upset, irritation, and loneliness, ranging from 0 to 27 (Cai et. al., 2022). Positive mood score is from questions relating to stress management, determination, and being alert, energetic, happy, inspired, and outgoing, with scores ranging from 0 to 28. Across participants, missing rates in mood scores vary from less than 10\% to over 90\%. We applied new methods SSMimpute, MLEEM, MLEN, MLENS, and compare them to existing methods of complete case analysis, LOCF, linear and spline interpolation, Kalman smoothing imputation, mean imputation, and MICE for unit root testing. We additionally considered MNAR senstivity analyses for the SSMimpute and MLEEM methods. As the mood score errors are not normally distributed, we found MLE methods conditioning on normality were unfit, and tended to over-report non-stationarity and estimate $\rho$ close to 1. To evaluate the performance of other methods, we highlight the time series of three participants. 

\begin{figure}
\centerline{%
\includegraphics[width = 170mm]{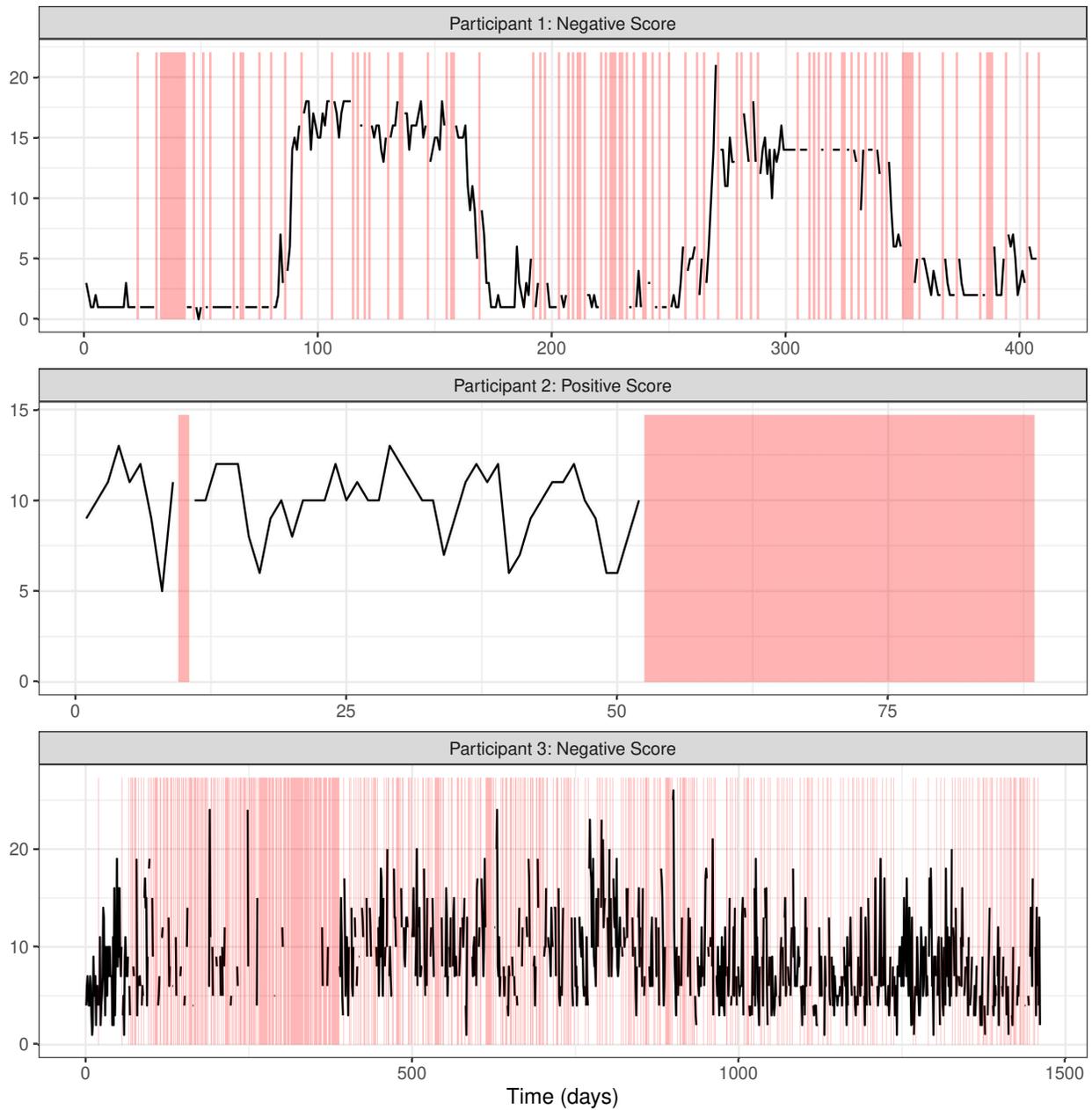}}
\caption{Time series plots for three participants included in BLS study. Missing observations are marked by red shading.}
\label{Figure 4}
\end{figure}

We first focus on the negative mood score for a participant with bipolar disorder followed for 708 days, with a 22.6\% missing rate in negative mood. We will concentrate on the first 407 days of follow up, prior to a significant life change for the patient which created a change point in the observed data. As seen in Figure 4, the negative mood time series of participant 1 appears non-stationary with potential seasonality. All methods but mean and MICE imputation had a p-value greater than 0.05, and thus failed to reject the null hypothesis of unit root. These results concord with those found in the simulation, where MICE and mean imputation strongly bias test results towards stationarity regardless of the true $\rho$. For reference, all but mean imputation and MICE estimated the autocorrelation to be greater than 0.95. As expected, a sensitivity analysis for the SSMimpute and MLEEM methods with $\delta$ values ranging from 0 to 1 had no impact on failing to reject the null of non-stationarity. 

We additionally consider the positive mood score of a participant with bipolar disorder with 89 days of follow up and a 41.6\% missing rate in positive mood observations. In Figure 4, we see for the positive score of participant 2 almost all missing time points are consecutive, with only one observation following the missing block.  All methods but spline interpolation, Kalman smoothing imputation, MLEN and MLENS reject the null of unit root. This reflects results seen in the simulation, where  spline interpolation and Kalman smoothing imputation appear to bias towards non-stationarity, reflected by an inflation in the type I error. Comparing autocorrelation estimates, those of linear and spline interpolation and Kalman smoothing imputation are larger than 0.6, while for complete case analysis, SSMimpute, and LOCF they are less that 0.4. This example of a short time series with high rates of missing data demonstrates the discrepancies in results that can be obtained using different methods, and the importance of considering the impact of the chosen method on autocorrelation estimation and unit root testing. When considering values of $\delta$ from 0 to 1 for a MNAR senstivity analysis, MLEEM with a $\delta \geq 0.5$ fails to reject the null hypothesis. SSMimpute shows increased estimates of autocorrelation, but across all considered $\delta$ concludes that the time series is stationary. 

Lastly, we include the negative mood score of a participant with schizophrenia and 1461 days of follow up with 34.6\% missing. In Figure 4, the time series visually appears stationary, with intermittent missing. Indeed, all methods reject the null hypothesis of unit root and find the autocorrelation to be less than one. This example demonstrates that with sufficiently low value of $\rho$, a relatively low missing rate, and a long follow up, the power is high enough that the chosen method to handle missingness has no impact on unit-root testing. Even with in sensitivity testing, we found little changes in autocorrelation estimates and no changes by Dickey Fuller test results across $\delta$ values.

\section{Discussion}
\label{s:discuss}

The increase in not fully observed time series data from contexts such as mHealth necessitates the adaptation of current time series analysis methods to a missing data setting. In particular, testing a univariate time series for stationarity is an important evaluation to inform appropriate subsequent analyses. However, there are few recommendations for testing unit root with missing data, and no consideration for missing mechanisms beyond missing completely at random. Here we propose adaptations to the ADF test including an MLE approach for testing for unit root when the time series has lag order of one and normal and independent errors, and a state space model multiple imputation method to test for unit root when these conditions are not met. We additionally introduce sensitivity analyses to evaluate the impact of data MNAR on test results and autocorrelation estimation. In our simulation we find that both methods are effective under MCAR and MAR mechanisms, with improved power and reasonable type I error when missingness is moderate. We additionally show that our proposed sensitivity analyses are effective at generating a range of autocorrelation estimates and test results depending on the missing mechanism assumptions. By applying the existing univariate time series missing methods and our proposed methods to mood time series from the Bipolar Longitudinal Study, we demonstrate that our simulation results hold in observed data, and the limitation of the proposed MLE method when errors are not normally distributed. 

The proposed SSMimpute, MLEEM, and MLENS methods offer computational efficiency, with high power and acceptable type I error across MCAR and MAR time series with varying rates of missingness. They additionally offer researchers the ability to evaluate the impact of hypothesized MNAR mechanisms on test results through sensitivity analyses. When noise is normally distributed and there is a single lag error, we recommend the MLENS method which demonstrates high power and low type I error when correctly specified. Should data be hypothesized to be missing not at random, the MLEEM method can be employed for sensitivity analyses. Should the time series follow a more complex generation process, the SSMimpute method is most appropriate, as it can handle non-normal noise and lag order greater than one. These methods will allow for improved validity of future mHealth analyses by accurate unit root testing to inform model selection and appropriate assumptions.  However, the general limitations in power of the ADF test, especially when autocorrelation is close to one or there are few observations (Harris, 1992) still apply to the proposed methods. Additionally, the ADF test and thus proposed adaptations to address missing data only test for unit root non-stationarity. Other violations to stationarity such as change points and unstable variance could also be found in mHealth data, and should also be tested for to inform subsequent analyses. Further research is needed to develop hypothesis tests for additional types of non-stationarity in the context of missing data. We also note that more research is needed to provide a framework for identifying the correct lag order under missing data.


\section*{Acknowledgements}

This work is supported by K01MH118477 (PI: Valeri) and U01MH116925 (PI: Baker). 





\section*{Data Availability Statement}
Data used in this paper to illustrate the proposed methods are not shared due to privacy restrictions.






\newpage


\begin{thebibliography}{}






\bibitem{}Aledavood, T., Hoyos, A. M. T., Alakörkkö, T., Kaski, K., Saramäki, J., Isometsä, E., \& Darst, R. K. (2017). Data collection for mental health studies through digital platforms: requirements and design of a prototype. JMIR research protocols, 6(6), e6919. 

\bibitem{}Azur, M. J., Stuart, E. A., Frangakis, C., \& Leaf, P. J. (2011). Multiple imputation by chained equations: what is it and how does it work?. International journal of methods in psychiatric research, 20(1), 40-49.

\bibitem{}Barnett, I., \& Onnela, J. P. (2020). Inferring mobility measures from GPS traces with missing data. Biostatistics, 21(2), e98-e112.

\bibitem{}Bertin, K., Torres, S., \& Tudor, C. A. (2011). Maximum-likelihood estimators and random walks in long memory models. Statistics, 45(4), 361-374.

\bibitem{}Cai, X., Wang, X., Eichi, H. R., Ongur, D., Dixon, L., Baker, J. T., ... \& Valeri, L. (2022). State space model multiple imputation for missing data in non-stationary multivariate time series with application in digital Psychiatry. arXiv preprint arXiv:2206.14343.

\bibitem[]{}Dickey, D. A., \& Fuller, W. A. (1979). Distribution of the estimators for autoregressive time series with a unit root. Journal of the American statistical association, 74(366a), 427-431. 

\bibitem{}Eekhout, I., Van De Wiel, M. A., \& Heymans, M. W. (2017). Methods for significance testing of categorical covariates in logistic regression models after multiple imputation: power and applicability analysis. BMC medical research methodology, 17(1), 1-12.

\bibitem{}Goldberg, S. B., Bolt, D. M., \& Davidson, R. J. (2021). Data Missing Not at Random in Mobile Health Research: Assessment of the Problem and a Case for Sensitivity Analyses. Journal of medical Internet research, 23(6), e26749.

\bibitem{}Harris, R. I. (1992). Testing for unit roots using the augmented Dickey-Fuller test: Some issues relating to the size, power and the lag structure of the test. Economics letters, 38(4), 381-386.

\bibitem{}Huang, E., \& Onnela, J. P. (2019). Activity Classification Using Smartphone Gyroscope and Accelerometer Data. arXiv preprint arXiv:1903.12616.

\bibitem{}Little, R. J., \& Rubin, D. B. (2019). Statistical analysis with missing data (Vol. 793). John Wiley \& Sons.

\bibitem{}Luckett, D. J., Laber, E. B., Kahkoska, A. R., Maahs, D. M., Mayer-Davis, E., \& Kosorok, M. R. (2019). Estimating dynamic treatment regimes in mobile health using v-learning. Journal of the American Statistical Association.

\bibitem{}Mandel, F., Ghosh, R. P., \& Barnett, I. (2021). Neural Networks for Clustered and Longitudinal Data Using Mixed Effects Models. Biometrics.

\bibitem{}Metcalfe, A. V., \& Cowpertwait, P. S. (2009). Introductory time series with R (p. 2). Springer-Verlag New York.

\bibitem{}Moritz, S., \& Bartz-Beielstein, T. (2017). imputeTS: time series missing value imputation in R. R J., 9(1), 207.

\bibitem{}Neto, E. C., Prentice, R. L., Bot, B. M., Kellen, M., Friend, S. H., Trister, A. D., ... \& Mangravite, L. (2016). Towards personalized causal inference of medication response in mobile health: an instrumental variable approach for randomized trials with imperfect compliance. arXiv preprint arXiv:1604.01055.


\bibitem{}Onnela, J. P., Dixon, C., Griffin, K., Jaenicke, T., Minowada, L., Esterkin, S., ... \& Jones, E. (2021). Beiwe: a data collection platform for high-throughput digital phenotyping. Journal of Open Source Software, 6(68), 3417.

\bibitem{}Park, S. J., Shin, D. W., Park, B. U., Kim, W. C., \& Oh, M. S. (2005). Bayesian test for asymmetry and nonstationarity in MTAR model with possibly incomplete data. Computational statistics \& data analysis, 49(4), 1192-1204.

\bibitem{}Silva, B. M., Rodrigues, J. J., de la Torre Díez, I., López-Coronado, M., \& Saleem, K. (2015). Mobile-health: A review of current state in 2015. Journal of biomedical informatics, 56, 265-272.

\bibitem[]{}Shin, D. W., \& Sarkar, S. (1996). Testing for a unit root in an AR (1) time series using irregularly observed data. Journal of Time Series Analysis, 17(3), 309-321.

\bibitem[]{}Shin, D. W., \& Sarkar, S. (1994). Unit root tests for ARIMA (0, 1, q) models with irregularly observed samples. Statistics \& Probability Letters, 19(3), 189-194.

\bibitem{}Skjelbred, H. I., \& Kong, J. (2019, March). A comparison of linear interpolation and spline interpolation for turbine efficiency curves in short-term hydropower scheduling problems. In IOP Conference Series: Earth and Environmental Science (Vol. 240, No. 4, p. 042011). IOP Publishing.

\bibitem{}Terry, W. R., Lee, J. B., \& Kumar, A. (1986). Time series analysis in acid rain modeling: Evaluation of filling missing values by linear interpolation. Atmospheric Environment (1967), 20(10), 1941-1943.

\bibitem{}Tewari, A., \& Murphy, S. A. (2017). From ads to interventions: Contextual bandits in mobile health. In Mobile Health (pp. 495-517). Springer, Cham.

\bibitem{}Torous, J., Staples, P., Barnett, I., Sandoval, L. R., Keshavan, M., \& Onnela, J. P. (2018). Characterizing the clinical relevance of digital phenotyping data quality with applications to a cohort with schizophrenia. NPJ digital medicine, 1(1), 1-9. 

\bibitem{} Torous, J., Kiang, M. V., Lorme, J., \& Onnela, J. P. (2016). New tools for new research in psychiatry: a scalable and customizable platform to empower data driven smartphone research. JMIR mental health, 3(2), e5165. 

\bibitem{}Vaidya, A. S., Srinivas, M. B., Himabindu, P., \& Jumaxanova, D. (2013, July). A smart phone/tablet based mobile health care system for developing countries. In 2013 35th Annual International Conference of the IEEE Engineering in Medicine and Biology Society (EMBC) (pp. 4642-4645). IEEE.

\bibitem{}van Buuren S, \& Groothuis-Oudshoorn K (2011). “mice: Multivariate Imputation by Chained Equations in R.” Journal of Statistical Software, 45(3), 1-67.

\bibitem{}van de Wiel, M. A., Berkhof, J., \& van Wieringen, W. N. (2009). Testing the prediction error difference between 2 predictors. Biostatistics, 10(3), 550-560.

\bibitem{}Wijesekara, W. M. L. K. N., \& Liyanage, L. (2020, March). Comparison of imputation methods for missing values in air pollution data: Case study on sydney air quality index. In Future of Information and Communication Conference (pp. 257-269). Springer, Cham.
 
\end{thebibliography}
\end{document}